\documentclass[%
aps,
]{revtex4-2}

\usepackage{graphicx}
\usepackage{dcolumn}
\usepackage{bm}
\usepackage{hyperref}
\usepackage{xcolor}

\begin{document}

\preprint{APS/123-QED}

\title{Finite speed of sound effects on asymmetry in multibubble cavitation}

\author{Mandeep Saini}
 \affiliation{Institut Jean le Rond $\partial$'Alembert CNRS UMR 7190,\\ Sorbonne Universit{\'e} \\
 F75005 Paris, France \\ Physics of Fluids Group, Max Planck Centre for Complex Fluid Dynamics, J.M. Burgers Centre for Fluid Dynamics, University of Twente, P.O. Box 217, 7500AE Enschede, The Netherlands}
 \email{mandeep.saini@sorbonne-universite.fr}
\author{Youssef Saade}%
 \email{y.saade@utwente.nl}
\affiliation{%
Physics of Fluids Group, Max Planck Centre for Complex Fluid Dynamics, J.M. Burgers Centre for Fluid Dynamics, University of Twente, P.O. Box 217, 7500AE Enschede, The Netherlands
}%

\author{Daniel Fuster}
 \affiliation{Institut Jean le Rond $\partial$'Alembert CNRS UMR 7190,\\ Sorbonne Universit{\'e} \\
 F75005 Paris, France}
\email{daniel.fuster@sorbonne-universite.fr}
\author{Detlef Lohse}
\affiliation{%
Physics of Fluids Group, Max Planck Centre for Complex Fluid Dynamics, J.M. Burgers Centre for Fluid Dynamics, University of Twente, P.O. Box 217, 7500AE Enschede, The Netherlands
}%

\affiliation{
Max Planck Institute for Dynamics and Self-Organisation, Am Fassberg 17, 37077 Göttingen, Germany
}
\email{d.lohse@utwente.nl}


\begin{abstract}
Three-dimensional direct numerical simulations (DNS) are used to revisit the experiments on multibubble cavitation performed by Bremond \textit{et al}. (\href{https://doi.org/10.1063/1.2396922}{Phys. Fluids 18, 121505 (2006)}, \href{https://doi.org/10.1103/PhysRevLett.96.224501}{Phys. Rev. Lett. 96, 224501 (2006)}). In particular, we aim at understanding the asymmetry observed therein during the expansion and collapse of bubble clusters subjected to a pressure pulse. Our numerical simulations suggest that the asymmetry is due to the force applied by the imposed pressure pulse and it is a consequence of the finite effective speed of sound in the liquid. By comparing our numerical results to the experiments, we found that the effective speed of sound under the experimental conditions was smaller than that of degassed water due to microbubbles in the system which resulted from prior cavitation experiments in the same setup. The estimated values of the effective speed of sound are consistent with those derived from the classical theory of wave propagation in liquids with small amounts of gas. To support this theory, we also present evidence of tiny bubbles remaining in the liquid bulk as a result of the fragmentation of large bubbles during the prior cavitation experiments. Furthermore, we find that this asymmetry also alters the direction of the liquid jet generated during the last stages of bubble collapse.

\end{abstract}

\maketitle


\section{\label{sec:intro}Introduction}
Cavitation bubbles are key components in a plethora of applications such as ultrasound imaging, high-intensity focused ultrasound treatment, drug delivery, lithotripsy etc. \cite{lohse2018bubble,brennen2015cavitation}. In many of these applications, multiple cavitation bubbles are exposed to pressure waves \cite{sapozhnikov2002effect,bremond2006interaction,PhysRevLett.96.224501,pishchalnikov2019high}. To achieve a well controlled process, it is important to understand the interaction among those bubbles, their surrounding medium, and the pressure waves that drive their motion. Several studies, have contributed to a better understanding and modeling of this intricate phenomenon \cite{fernandez2012localized,stricker2013interacting,fuster2011modelling,fuster2019review,fan2021time}. Bremond \textit{et al}. \cite{bremond2006interaction,PhysRevLett.96.224501} conducted comprehensive and well controlled experiments by fixing the bubble locations using micropits, which allowed them to study the dynamics of a bubble cluster, and its interaction with the pressure pulse. However, the numerical modeling therein was limited to a modified Rayleigh-Plesset model and an axisymmetric boundary integral method due to the very high computational costs of three-dimensional direct numerical simulations. In this study, we numerically revisit the multibubble cavitation problem from Refs. \cite{bremond2006interaction,PhysRevLett.96.224501} using three-dimensional, compressible flow, direct numerical simulations, with the volume of fluid (VOF) method. The objective is to study the breaking of spherical symmetry and the compressibility effects emerging from the interaction between the bubbles and the pressure pulse. 

\begin{figure}
    \centering
    \includegraphics[]{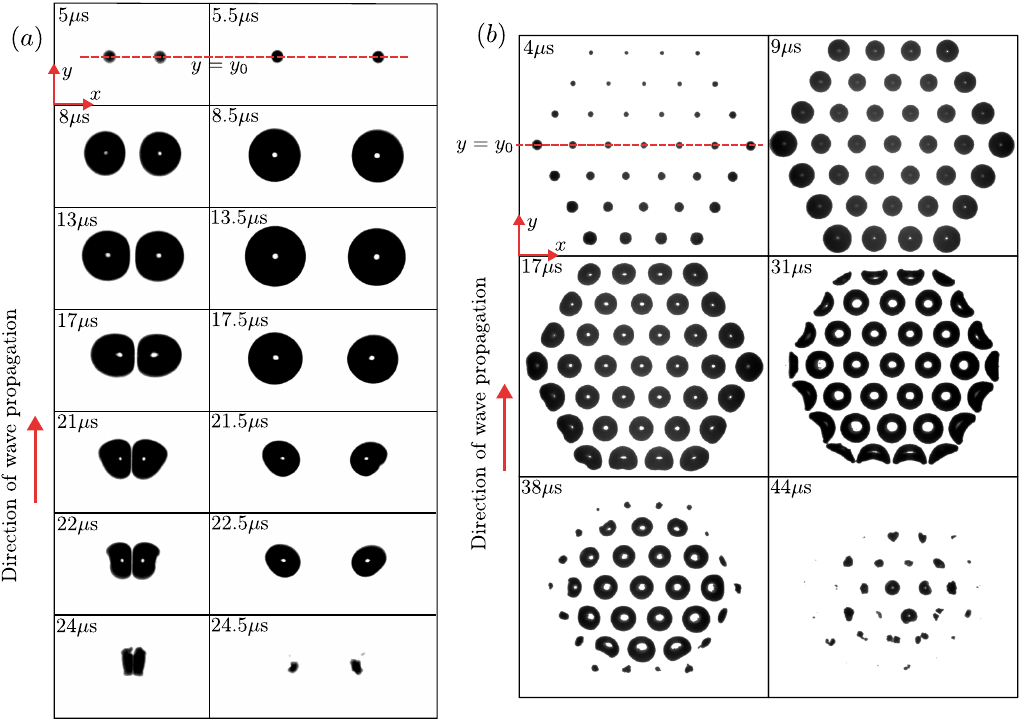}
    \caption{Top view of multiple cavitation bubbles nucleating from pits ($4$ $\mu {\textrm m}$ diameter) and collapsing in contact with a rigid wall. $(a)$ A pair of bubbles nucleating from two pits separated by distances of $200$ and $400$ $\mu {\textrm m}$ are shown in the left and right panels, respectively. $(b)$ A cluster of 37 bubbles nucleating from the pits drilled in a hexagonal configuration having a pitch distance of $200$ $\mu {\textrm m}$. This figure is adapted from the experiments of Bremond \textit{et al}. (2006) \cite{PhysRevLett.96.224501,bremond2006interaction}.}
    \label{fig:asymexpt}
\end{figure}

Figure \ref{fig:asymexpt} shows snapshots taken from the experiments of Bremond \textit{et al.} \cite{bremond2006interaction,PhysRevLett.96.224501} in which 4 $\mu \textrm{m}$ pits were etched on a silicon plate and were later submerged in a large water tank.  A pulse generated by a piezo transducer caused a pressure drop, leading to the growth of the nuclei in the pits into large bubbles up to 100 $\mu \textrm{m}$ in size. The top views from three cases are shown: two pits drilled at a distance of 200 $\mu \textrm{m}$ [left panel of Fig. \ref{fig:asymexpt}$(a)$], two pits separated by a distance of 400 $\mu \textrm{m}$ [right panel of Fig. \ref{fig:asymexpt}$(a)$], and 37 pits etched in a hexagonal arrangement of 200 $\mu \textrm{m}$ pitch [Fig. \ref{fig:asymexpt}$(b)$]. The symmetry breaking effects in the direction of wave propagation can be observed at 22--24 $\mu \textrm{s}$ in the bubble pair case. In the absence of compressibility effects, the bubble dynamics would exhibit mirror symmetry about the $y = y_0$ line. However, this symmetry is broken in the experiments since water is slightly compressible. We speculate that this symmetry breaking could be a result of the force induced by the pressure gradient due to the applied pressure pulse and/or due to a Rayleigh--Taylor instability. It is well known that when a single bubble in the liquid bulk is exposed to a pressure wave, this wave induces a force that is responsible for the translational motion of the bubble \cite{prosperetti1984bubble,bjerknes1906fields,leighton1990primary,parlitz1999spatio,leighton2012acoustic,doinikov2005bjerknes}. The instantaneous force induced by pressure gradient $\boldsymbol{\nabla} p (\mathbf{x},t)$ on a bubble of volume $V(t)$ is  
\begin{equation}
    \mathbf{F} (t) \propto - V(t)\boldsymbol{\nabla} p (\mathbf{x},t).
\end{equation}
\noindent If the bubble is attached to a wall, the viscous dissipation hinders the motion of the contact line. Therefore, this force can break the spherical symmetry of the bubble. For the multibubble case, the neighboring bubbles can also alter the pressure field, thus inducing pressure gradients, increasing the asymmetry further. On the other hand, small non-spherical perturbations in the bubble shape can grow unstable during the deceleration phase of the collapsing bubble, which is known as Rayleigh-Taylor (RT) instability \cite{plesset1956stability,brennen2014}. Thus, we postulate that the asymmetry is initiated during the expansion phase due to the pressure gradient and that this asymmetry can get amplified during the collapse phase due to RT instabilities.

This article is organized as follows: Sec. \ref{sec:method} outlines the governing equations and the problem setup. Sec. \ref{sec:pair} delves into the dynamics of bubble pairs and provides a detailed investigation of the asymmetrical bubble response. Sec. \ref{sec:multi} presents the results for the inception and collapse of arrays of more than two bubbles. Finally, we draw conclusions in Sec. \ref{sec:conclusion} and present an outlook for future studies.


\section{\label{sec:method}Method}
We use a compressible all-Mach solver that was recently developed and implemented in the free software program Basilisk \cite{basilisk,popinet2015quadtree,fuster2018all}. The software is equipped with octree-based mesh refinement capabilities which are essential in performing three-dimensional simulations. This all-Mach solver has been extensively used to study various bubble dynamics problems \cite{fan2020optimal,saade2021crown,li2021comparison,saini2022dynamics,saade2023multigrid}. The governing equations are the conservation of mass, momentum, and energy. In the absence of heat and mass transfer, these equations can be written for a two-phase flow as

\begin{eqnarray}
        \frac{\partial \rho_i}{\partial t} + \boldsymbol{\nabla} \boldsymbol{\cdot} (\rho_i {\mathbf u}_i) & = & 0, \label{eq:dnsrho}\\
        \frac{\partial (\rho_i {\mathbf u}_i)}{\partial t} + \boldsymbol{\nabla} \boldsymbol{\cdot} (\rho_i {\mathbf u}_i{\mathbf u}_i) & = & -\boldsymbol{\nabla} p_i + \boldsymbol{\nabla} \boldsymbol{\cdot} {\boldsymbol{\tau}}_i, \label{eq:dnsmom}\\
        \frac{\partial (\rho_i E_i)}{\partial t} + \boldsymbol{\nabla} \boldsymbol{\cdot} (\rho_i E_i \mathbf{u}_i) & = & - \boldsymbol{\nabla} \boldsymbol{\cdot} (p_i {\mathbf u}_i ) + \boldsymbol{\nabla} \boldsymbol{\cdot} ({\boldsymbol{\tau}}_i \mathbf{\cdot} {\mathbf u}_i), \label{eq:dnsenergy}
\end{eqnarray}

\noindent where the subscript $i \in \{l,g\}$ denotes the liquid and gas phases, respectively; $\rho_i$ is the density; $\mathbf{u}_i$ the velocity vector field; $p_i$ the pressure field; $E_i$ the total energy per unit volume, defined as the sum of the internal and kinetic energies $E_i = e_i + \frac{1}{2} \mathbf{u}_i^2$; and $\boldsymbol{\tau}_i = \mu_i (\boldsymbol{\nabla} {\mathbf u}_i + (\boldsymbol{\nabla} {\mathbf u}_i)^{\textrm T} - \frac{2}{3} \boldsymbol{\nabla} \boldsymbol{\cdot} \mathbf{u_i} {I})$ the viscous stress tensor, with $I$ being the identity tensor. The system of equations is then closed by the stiffened gas equation of state (EOS), defined as

\begin{equation}
    \rho_i e_i = \frac{p_i + \Gamma_i \Pi_i}{\Gamma_i - 1},
    \label{eq:EOS}
\end{equation}

\noindent where $\Gamma_i$ and $\Pi_i$ are empirical parameters obtained by fitting the speed of sound. For water $\Gamma_l = 5.5$ and $\Pi_l = 4921$ bars \cite{johnsen2006implementation}. In Sec. \ref{sec:pair}, we vary the parameter $\Pi$ to change the effective speed of sound in the medium. The speed of sound $c_i$ in the medium is defined by the EOS as 

\begin{equation}
    c_i = \sqrt{\Gamma_i \frac{p_i + \Pi_i}{\rho_i}}.
\end{equation}
For an ideal gas, $\Gamma_g = \gamma$ is the ratio of specific heats, and $\Pi_g = 0$. 

The interface conditions couple the motion of fluids in each phase. In the absence of mass transfer the velocity is continuous across the interface such that $[[\mathbf{u}]] = 0$, where $[[\cdot]]$ represents the jump in the particular quantity across the interface. The pressures in both phases are related by the Laplace equation $[[p]] = - \sigma \kappa + [[ \mathbf{n}_I \boldsymbol{\cdot} \boldsymbol{\tau} \boldsymbol{\cdot} \mathbf{n}_I ]]$,
where $\sigma$ is the surface tension coefficient, $\kappa$ is the curvature, and $\mathbf{n}_I$ is the unit vector normal to the interface. We also assume that no heat is transferred across the interface, which implies that the normal derivative of the internal energy remains continuous, i.e., $[[\partial e/\partial n]] = 0.$

\begin{figure}[h!]
    \centering
    \includegraphics[]{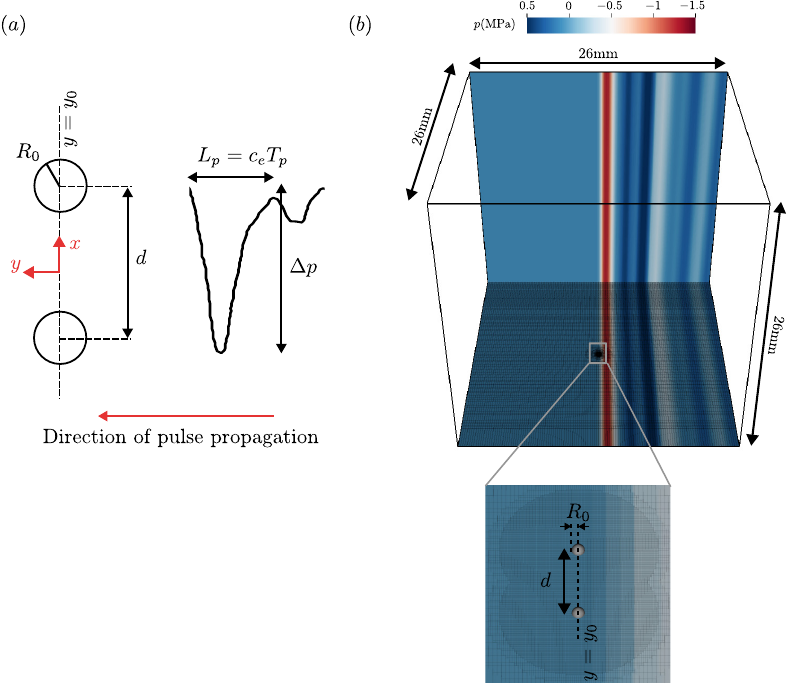}
    \caption{(a) The simplified schematic (top view) of the problem where the initial nuclei (shown with circles) of radii $R_{0}$ are separated by a distance $d$ and exposed to a 1D pressure pulse of amplitude $\Delta p = 1.5 \textrm{MPa}$ and a characteristic length $(L_p)$ (defined from the pulse duration $T_p$ given from hydrophone measurements). The direction of wave propagation is shown by the red arrow. (b) The 3D computational setup used in the current study where the projected color map corresponds to the initial pressure field taken from the experiments \cite{PhysRevLett.96.224501}. A zoomed-in view of the initial nuclei and the projection of the numerical grid on the bottom boundary of the domain are shown.}
    \label{fig:setupprob}
\end{figure}

The above equations are discretized using a finite volume method while satisfying the interface conditions. We use a geometric VOF method to track the interface between the two fluids \cite{tryggvason2011direct}. In the VoF method, the phase characteristic function is represented by the color function $C_i$ in the discrete cells. It is equal to 1 in the reference phase, to 0 in the non-reference phase, and to a fractional value between 0 and 1 in the cells containing both liquid and gas phases. The conserved quantities (density $C_i \rho_i$, momentum $C_i \rho_i \mathbf{u}_i$, total energy $C_i \rho_i E_i$) are advected consistently with the color function (see Ref. \cite{arrufat2021mass}). The interface is reconstructed at each time step using piecewise-linear constructions. Besides this, we use a one fluid formulation to solve the equation for the average momentum 

\begin{equation}
\frac{\partial \overline{\rho \mathbf{u}}}{\partial t} + \boldsymbol{\nabla \cdot} (\overline{\rho \mathbf{u}} \textrm{ } \overline{\mathbf{u}}) = - \boldsymbol{\nabla} \overline{p} + \boldsymbol{\nabla \cdot} {\overline{\boldsymbol{\tau}}} + \sigma \kappa \delta_s \mathbf{n}_I,   
\end{equation}

\noindent where $\overline{\Phi}$ represents the average value of a particular quantity $\Phi$, defined as $\overline{\Phi} = C \Phi_1 + (1 - C) \Phi_2$; $\delta_s$ is the delta function; and $\mathbf{n}_I$ is the unit vector normal to the interface. The capillary forces are added as continuum surface forces, where $\delta_s$ is approximated as the gradient of the color function $\vert \boldsymbol{\nabla} C \vert$ \cite{brackbill1992continuum}. This method is well balanced and ensures momentum conservation \cite{rudman1998volume}. For more details about the numerical method, the reader is referred to Fuster and Popinet \cite{fuster2018all}. The motion of the contact line is regularized by the Navier-slip model, 
\begin{equation}
\mathbf{u} - (\mathbf{u} \boldsymbol{\cdot} \mathbf{n})\mathbf{n} = \lambda_{num} \left( \mathbf{\boldsymbol{\tau \cdot} n} - \left[(\mathbf{\boldsymbol{\tau \cdot} n}) \mathbf{\boldsymbol{\cdot} n}\right] \mathbf{n} \right)
\end{equation}

\noindent where $\mathbf{u}$ is the velocity vector at the wall, $\boldsymbol{\tau}$ is the stress tensor at the wall, $\mathbf{n}$ is the unit vector normal to the wall, and $\lambda_{num}$ is the slip length. In Appendix, we show that our results are independent of the contact line dynamics. We also use a static contact angle model implemented using the approach of Afkhami and Bussmann \cite{afkhami2009}. For other boundaries, we use reflecting boundary conditions, i.e., $\mathbf{u \boldsymbol{\cdot} n} = 0$, $\boldsymbol{\nabla \cdot} \left[\mathbf{u} - (\mathbf{u \boldsymbol{\cdot} n}) \mathbf{n}\right] = 0$, and $\mathbf{n \boldsymbol{\cdot} } \boldsymbol{\nabla} p = 0$. The numerical domain is made large enough (see Fig. \ref{fig:setupprob}) to prevent any reflected pressure waves from influencing the bubble dynamics.

A simplified schematic (top view) of the problem is depicted in Fig. \ref{fig:setupprob}$(a)$. The initial nuclei, represented as circles of radii $R_{0}$, are separated by a distance $d$ and exposed to a one-dimensional (1D) pressure pulse of amplitude $\Delta p \approx 1.5 
\textrm{ MPa}$ that propagates with an effective speed of sound towards the nuclei, along the $y$ axis. This pulse is applied for constant duration $T_p$; hence, one can readily define a characteristic length for the pulse as $L_p = c_e T_p$, where $c_e$ is the effective speed of sound in the liquid. Note that the geometry exhibits mirroring symmetry about the $y$ axis.

Figure \ref{fig:setupprob}$(b)$ shows the computational setup. We use a cubic domain of size varying from $26\textrm{ mm}$ in the case of bubble pairs to $130 \textrm{ mm}$ in the case of a bubble cluster. The color map shows the projection of the initial pressure field which is taken from the hydrophone measurements of Bremond \textit{et al}. \cite{bremond2006interaction}. The temporal pressure perturbation $p^{\prime} (t)$ at $y_0$ is transformed into the spatial domain $(y = y_0 + c_e t)$ using $p(y_0 + c_e t) = p_0 + p^\prime (y_0 + c_e t)$. Initial density and velocity perturbations are calculated from $p^\prime$ as $\rho^\prime = p^\prime/c_e^2$ and $u_y^\prime = p^\prime/\rho c_e$ using linear wave propagation theory. The bottom boundary is considered a solid wall where we place the hemispherical nuclei of radii $R_{0}$. For the particular case of the bubble pair, a zoomed-in view of the bottom boundary is also given, showing the initial nuclei and the projection of the grid on the bottom boundary. The grid is progressively refined from $40$ $\mu {\textrm m}$ far from the bubbles to the smallest grid size of $5$ $\mu {\textrm m}$ near the nuclei. The mesh refinement capabilities of Basilisk allow us to resolve the large scale separation from the characteristic length $L_p$ of the pressure pulse to the bubble radius (up to 3 orders of magnitude). The simulations were performed on the Swiss super computer \emph{Piz Daint} on which a typical three-dimensional (3D) simulation takes around 15h on 7200 processors and solves for approximately 1.3 million grid points.

\section{\label{sec:pair}Bubble pair}

\begin{figure}[h!]
    \centering
    \includegraphics{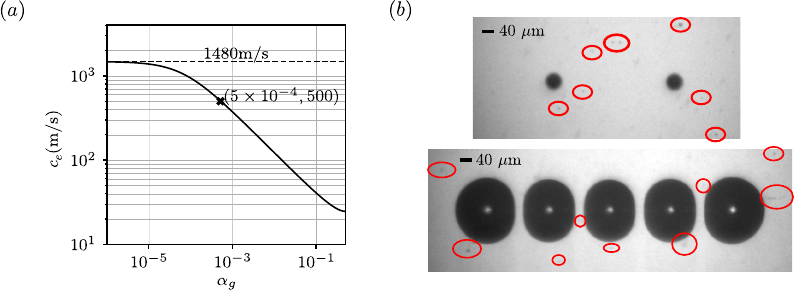}
    \caption{(a) The dramatic decrease in the effective speed of sound for a mixture of liquid and dispersed gas is shown as a function of the volume fraction of the gas phase $\alpha_g$ given by Eq. (\ref{eq:cavg}). The point obtained by matching $\mathcal{A}_y$ from experiments and numerical simulations is shown by a black cross, and its coordinates are specified in parentheses. (b) The tiny fragmented bubbles seen during the cavitation experiments are highlighted with red circles along with the bigger bubbles nucleating from the pits for the cases of two pits (top) and five pits (bottom). The scale in the top left corners of the top and bottom panels is $40$ $\mu \textrm{m}.$}
    \label{fig:ClAll}
\end{figure}

In this section, we investigate the problem of a pair of bubbles separated by a distance $d$. Our aim is to understand the effect of $c_e$ and $R_{0}$ on the dynamic response of bubbles to the driving pressure pulse. The speed of sound in pure water is 1480 m/s; however, even a tiny concentration of small bubbles can drastically lower this value, thus slowing down the propagation of pressure waves in the liquid. Henceforth, we assume a homogeneous and isotropic distribution of monodispersed bubbles, enabling us to represent the system as an equivalent medium in which the effective speed of sound $c_e$ is given as a function of the volume fraction of the dispersed bubbly phase $\alpha_g$ \cite{wood1956textbook,brennen2014},

\begin{equation}
    \frac{1}{c_e^2} = \left[ \alpha_g c_g + (1 - \alpha_g) c_l \right] \left[\frac{\alpha_g}{\rho_g c_g^2} + \frac{1 - \alpha_g}{\rho_l c_l^2}\right],
    \label{eq:cavg}
\end{equation} 

\noindent where $c_l$ and $c_g$ are the speeds of sound in pure liquid and in the gas phase, respectively. In Fig. \ref{fig:ClAll}$(a)$, Eq. (\ref{eq:cavg}) is plotted on a log-log scale, showing that a tiny volume fraction ($\alpha_g \sim 10^{-3}$) of dispersed microbubbles can decrease the speed of wave propagation to a few hundreds of meters per second.
These microbubbles can nucleate due to rarefaction waves and/or can also result from the bubble fragmentation during a series of experiments carried out at very short time intervals ($\sim 1$ $\mu s$). It must be stated that these experiments were rapidly repeated to capture the images of bubbles from the consecutive experiments using the stroboscopy technique (owing to the high reproducibility). Additionally in Fig \ref{fig:ClAll}$(b)$, we show the presence of extremely small bubbles of the order of $1$ $\mu \textrm{m}$ in the liquid. In numerical simulations, we model the effect of these small $1$ $\mu \textrm{m}$ bubbles by assuming a lower effective speed of sound compared to that of pure water in the bulk phase.

Equation (\ref{eq:cavg}) is applicable when the pulse has a small amplitude and a characteristic frequency much smaller than the bubble's resonance frequency ($\omega_p/\omega_b << 1$). In the experiments, the size of the fragmented small bubbles is of the order of $1$ $\mu \textrm{m}$, and the characteristic time scale of the pressure pulse (duration of negative pressure) is approximately $5$ $ \mu s$, resulting in a small frequency ratio $\omega_p/\omega_b \approx 0.037$. It is also important to note that the theory assumes a homogeneous dispersion of gas in liquid, while in experiments, the gas concentration can vary locally. Both non-linear effects and concentration gradients will certainly introduce some uncertainties in the simplified model used here. Nonetheless, we can safely assume that the experiments were done in a regime where the results were significantly influenced by the presence of very small amounts of the gas in the bulk phase and the resulting microbubbles due to the experimental procedure of repetitively sending the rarefaction waves through the liquid. 

\begin{figure}[b!]
    \centering
    \includegraphics[]{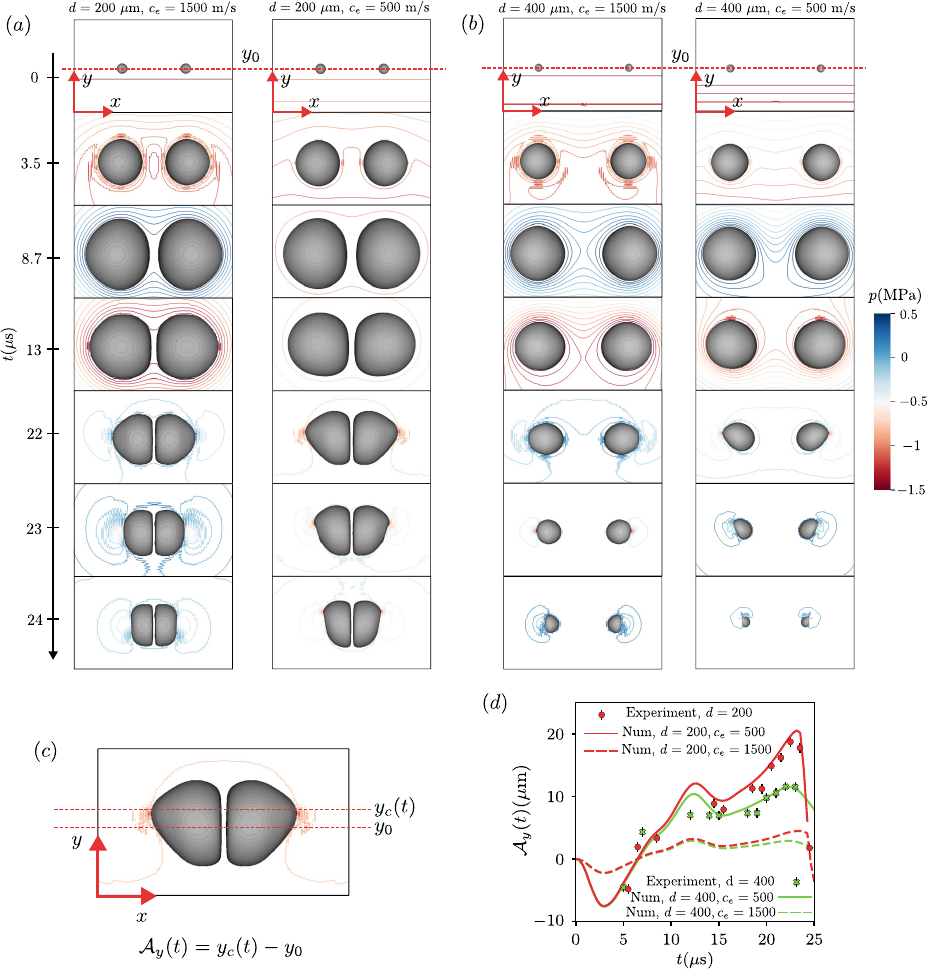}
    \caption{Comparison of the numerical results for two different values of the effective speed of sound, i.e. $c_e = 1500 \textrm{m/s}$ and $c_e = 500 {\textrm m/s}$. The bubble pair evolves from hemispherical nuclei with radii of $20$ $\mu {\textrm m}$. In $(a)$ and $(b)$, the top view of the bubble shape and pressure field are displayed at subsequent times shown. $(a)$ Separation distance $d = 200$ $\mu{\textrm m}$ between the nuclei with $c_e =1500 \textrm{ m/s}$ on the left and $c_e = 500 \textrm{ m/s}$ on the right. $(b)$ Separation distance $d = 400$ $\mu{\textrm m}$ between the nuclei with $c_e = 1500 \textrm{ m/s}$ on left and $c_e=500 \textrm{ m/s}$ on right. $(c)$ A particular snapshot showing the definition of the asymmetry parameter $\mathcal{A}_y(t)$, which is defined as the shift of the bubble centroid in the $y$ direction. $(d)$ The time evolution of $\mathcal{A}_y(t)$ obtained from the numerical simulations for all the cases discussed in $(a)$ and $(b)$ along with a comparison with experimental data. The error bars are equal to one pixel size ($1.53$ $\mu {\textrm m}$) in the experiment snapshots.}
    \label{fig:asymNum}
\end{figure}

We will begin by discussing the numerical results for four different cases obtained with combinations of pit distances $d \in \{200,400\}$ $\mu\textrm{m}$ and effective speeds of sound $c_e \in \{500,1500\} \textrm{ m/s}$. 
The radii of the nuclei in these simulations are fixed to $20$ $\mu {\textrm m}$. In Figs. \ref{fig:asymNum}$(a)$ and \ref{fig:asymNum}$(b)$, we show top view snapshots of the bubbles at various times for $d = 200$ $\mu \textrm{m}$ and $400$ $\mu \textrm{m}$, respectively. The left column in each panel corresponds to $c_e = 1500 \textrm{ m/s}$, while the right column corresponds to $c_e = 500 \textrm{ m/s}$. The isobars are depicted with color contours. The pressure pulse travels in the $y$ direction, resulting in a pressure drop that causes the rapid expansion of the bubbles, and their subsequent collapse at the end of the pulse. For both $d = 200$ $\mu \textrm{m}$ and $400$ $\mu \textrm{m}$, the bubble shapes observed in experiments (Fig. \ref{fig:asymexpt}) are better represented by the numerical results for $c_e = 500 \textrm{ m/s}$, especially during the collapse stage, when the bubbles exhibit an asymmetric behavior that is insignificant for $c_e = 1500 \textrm{ m/s}$.

In order to quantify the asymmetric response of the bubbles, we define the asymmetry parameter $(\mathcal{A}_y)$ as the shift of the bubble's centroid in the direction of motion of the pressure pulse, such that $\mathcal{A}_y(t) = y_c(t) - y_0$, where $y_c(t)$ is the bubble's centroid at an instant $t$ (see Fig. \ref{fig:asymNum}$(c)$). 
The numerical results (lines) shown in Fig. \ref{fig:asymNum}$(d)$ reveal that when the wave approaches the bubbles, the latter experience a negative pressure gradient and hence a force towards the negative $y$ direction, and thus negative values of $\textrm{A}_y$. Successively, as the pressure rises back to ambient values, the gradient of pressure becomes positive, and hence, the direction of the force applied by the wave on the bubbles and $\mathcal{A}_y$ change signs. Clearly, the experimentally observed asymmetric response (dots) of the bubbles is better captured numerically by $c_e = 500 \textrm{ m/s}$. 

We will now explore in detail the influence of the speed of sound and the initial bubble radii on the asymmetry in the following sections.

\subsection{Effect of the speed of sound}

\begin{figure}[b!]
    \centering
    \includegraphics[scale = 1]{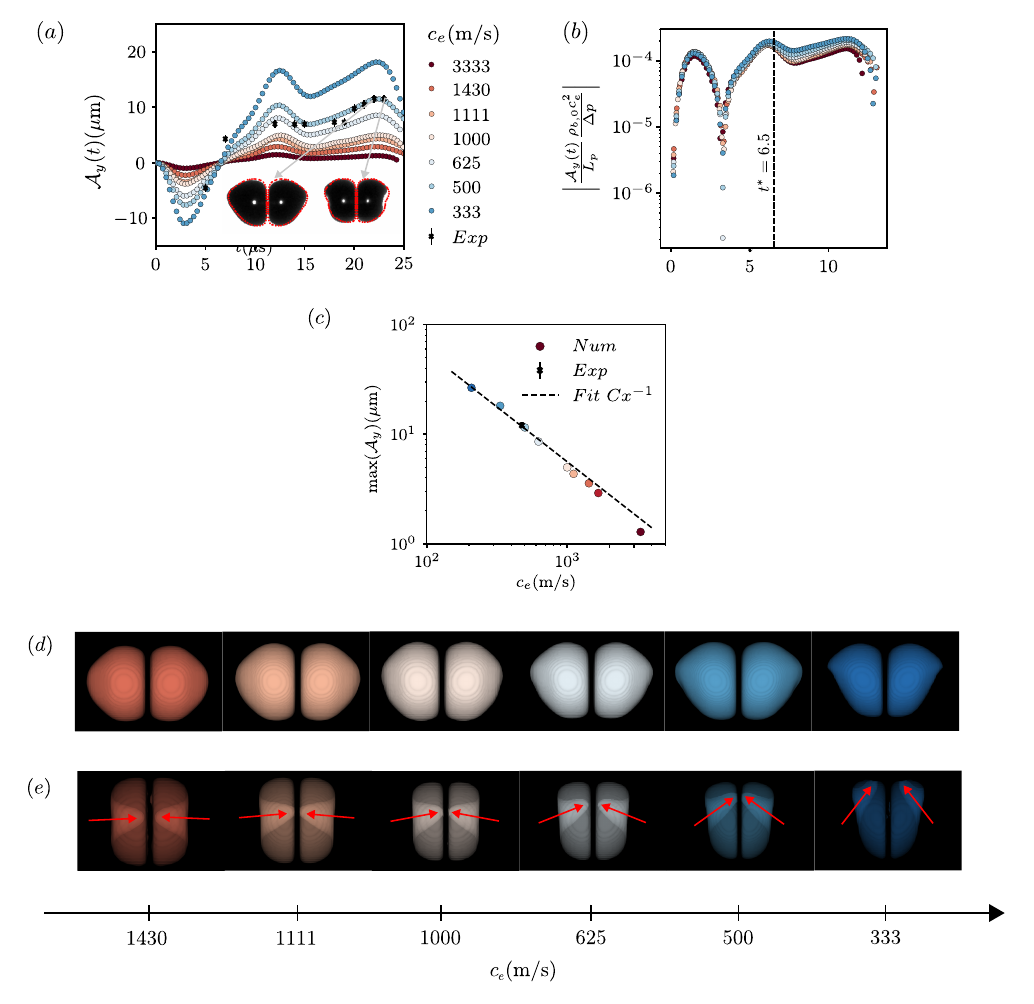}
    \caption{Effect of the effective speed of sound on the asymmetry for the particular case where the bubbles are nucleating from a pair of hemispherical nuclei with radii of $20$ $\mu {\textrm m}$ each, and separated by a distance $d=200$ $\mu {\textrm m}$. $(a)$ The time evolution of $\mathcal{A}_y$ is shown for different values of the effective speed of sound (colormap), alongside the experimentally obtained values (black crosses). The experimental bubble shapes overlaid with their numerical counterparts (red dots) at $t = 22$ and $23$ $\mu {\textrm s}$ are illustrated in the inset. $(b)$ The evolution of the dimensionless $\mathcal{A}_y(t)$ as a function of dimensionless time ($t^* = tU_c/R_0$) obtained using the scaling predicted from Eq. (\ref{eq:asymscaling}) for different values of $c_e$ $(c)$ The maximum value of the asymmetry parameter $\textrm{max}(\mathcal{A}_{y})$ is plotted on a log-log scale as function of the effective speed of sound along with the experimental point where its $c_e$ is obtained by matching $\textrm{max}(\mathcal{A}_{y})$ with the numerical data. A fit of all data (dashed line) is also plotted to demonstrate that $\textrm{max}(\mathcal{A}_{y})$ decays like $c_e^{-1}$ as predicted by Eq. (\ref{eq:asymscaling}) for given $T_p$. 
    $(d)$ The bubble shapes at the instant of $\textrm{max}(\mathcal{A}_{y})$ are shown for different effective speeds of sound in the liquid. $(e)$ Semitransparent bubble shapes are shown to visualize the jet generated during the last stages of collapse, whose direction is highlighted with red arrows.}
    \label{fig:asymMa}
\end{figure}

We consider the particular case of a pair of hemispherical nuclei with radii $R_0 = 20$ $\mu \textrm{m}$, separated by a distance $d = 200$ $\mu \textrm{m}$, and vary $c_{e}$ from 208 to 3333 \textrm{m/s}. Then, we analyze the temporal evolution of the asymmetry parameter $\mathcal{A}_y(t)$ for different values of $c_e$ (see Fig. \ref{fig:asymMa}$(a)$). The evolution of $\mathcal{A}_y(t)$ is similar in all the cases and varies as described before. The experimental measurement of $\mathcal{A}_y (t)$ (black crosses) is also plotted to compare with our numerical results. It is found that $c_e \approx 500\textrm{ m/s}$ best reproduces the asymmetry in the experiments, which had motivated the choice of this value of $c_e$ in figure \ref{fig:asymNum}. The inset of figure \ref{fig:asymMa}$a$ shows a comparison between the numerical (red points) and experimental bubble shapes at $t = 22$ and $23$ $\mu \textrm{s}$, confirming that the choice is a good fit.

As discussed in the Introduction (Sec. \ref{sec:intro}), during the expansion phase, the asymmetry can be the result of force induced by a pressure pulse on the bubbles or by each bubble on its neighbor. But since the problem is symmetric in the $y-z$ plane, the pressure gradient due to the neighboring bubble, acting in the $x$ direction, is expected to have a negligible effect on $\mathcal{A}_y$. Therefore, we postulate that the force due to the finite length scale of the pressure pulse $L_p$ is the main asymmetry mechanism during the expansion process. The component of this force in the $y$ direction $F_{y}$ at a given instant is

\begin{equation}
    F_{y}(t) \simeq - V_b \frac{\partial p}{\partial y} \approx - V_b \frac{\Delta p}{L_p},
    \label{eq:fby}
\end{equation}

\noindent where $V_b$ is the bubble volume. In writing this expression, we assume that the pressure induced by the pulse varies linearly in the $y$ direction. Equation (\ref{eq:fby}) therefore states that a reduction in the pulse length $L_p$ ($\propto c_e$) yields stronger pressure gradients and thus a stronger force. This leads to a more pronounced asymmetry, as observed in Fig. \ref{fig:asymMa}$(a)$. We assume that the motion of the bubble centroid scales with the force $F_y$ as 

\begin{equation} 
F_{y}(t) \sim \frac{\rho_{b,0} V_{b,0} \mathcal{A}_y (t)}{T_p^2},
\label{eq:scalingA}
\end{equation}

\noindent which can be interpreted as Newton's second law for the motion of the bubble centroid. We obtain from Eqs. (\ref{eq:fby}) and (\ref{eq:scalingA}) that the asymmetry parameter scales as

\begin{equation}
    \mathcal{A}_{y} (t) \sim - \frac{L_p}{\rho_{b,0}} \frac{\Delta p}{c_e^2} 
    \sim - \frac{T_p}{\rho_{b,0}} \frac{\Delta p}{c_e}.
    \label{eq:asymscaling}
\end{equation}

In Fig. \ref{fig:asymMa}$(b)$, we plot the evolution of nondimensional $\mathcal{A}_y$, obtained using the scaling given by Eq. (\ref{eq:asymscaling}), as a function of the dimensionless time $t^* = t U_c/R_0$ where $U_c = \sqrt{\Delta p/\rho_l}$. The curves for the different values of $c_e$ nicely collapse, pointing out the influence of the pressure gradient on the development of the asymmetry, especially during the expansion phase ($t^* \lesssim 6.5$). Some differences are detectable during the collapse phase $(t^* \gtrsim 6.5)$ that can be attributed to the RT instability which is known to grow when the liquid decelerates during the collapse phase ($\dot{R} < 0$, $\ddot{R} > 0$). Figure \ref{fig:asymMa}$(c)$ shows that the maximum value of the asymmetry $\textrm{max}(\mathcal{A}_y)$ also varies inversely with the effective speed of sound ($\propto c_e^{-1}$), similar to the instantaneous $\mathcal{A}_y(t)$ predicted by Eq. (\ref{eq:asymscaling}). 
The bubble shapes at the instant of maximum asymmetry ($t \approx 11.5$ $\mu \textrm{s}$) are also shown in Fig. \ref{fig:asymMa}$(c)$ for different values of $c_e$. 

We now discuss the effect of the asymmetry on the well-known jets which form during the collapse of the interacting bubbles \cite{tomita1990dynamic,bremond2006interaction,supponen2015inner}. Indeed, during the last stages of the collapse, the bubble pair forms liquid jets parallel to the wall, as shown in Fig. \ref{fig:asymMa}$(e)$. The direction of these jets is a function of the asymmetry, as highlighted with the arrows. At sufficiently small values of $\mathcal{A}_y$, the jets are directed along the $x$ direction. However, for large values of $\mathcal{A}_y$, the jets shift towards the direction of the propagation of the pressure pulse ($y$ axis). The jets themselves were not directly visible in the experiments of Bremond et al. \cite{bremond2006interaction,PhysRevLett.96.224501}; however, the observed kinks in the bubble interface in the experimental images [for instance, the left panel of Fig. \ref{fig:asymexpt}$(a)$ at 22s] serve as indicators of the presence of these jets in experiments.

\subsection{Effect of the nuclei radii}

\begin{figure}[h!]
    \centering
    \includegraphics{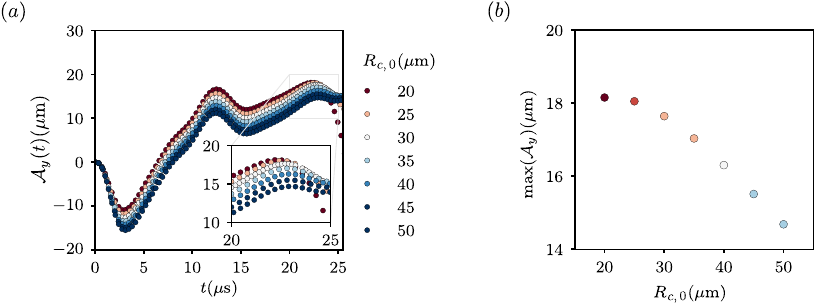}
    \caption{The numerical results for the particular case where the effective speed of sound is assumed to be $333{\textrm{ m/s}}$ and $d = 200$ $\mu \textrm{m}$ for varying $R_0$. $(a)$ The time evolution of the asymmetry parameter $\mathcal{A}_y(t)$ is shown for different sizes of  nuclei depicted with the color map. A zoomed-in view around $\textrm{max}(\mathcal{A}_y)$ is shown in the inset. $(b)$ The value of $\textrm{max} (\mathcal{A}_y)$ corresponding to the maximum in $(a)$ is plotted as a function of the sizes of the nuclei. The color code in $(a)$ and $(b)$ characterizes the radii of nuclei $R_0$}
    \label{fig:AsymSize}
\end{figure}
The experiments discussed by Bremond \textit{et al}. \cite{bremond2006interaction} were performed for cylindrical pits, yet the bubble evolution could be adequately predicted from a Rayleigh-Plesset model assuming hemispherically shaped nuclei. The radii $R_0$ of these nuclei were calculated from the pit diameter $D$ and height $H$ as $R_0 = (3HD^2/8)^{1/3}$, which gives $R_0 \approx 5 \textrm{ } \mu \textrm{m}$. However, in the current study, we could not initialize nuclei smaller than $20$ $\mu \textrm{m}$ in radius due to limited computational resources. For a single bubble, changing the initial radius from $5$ to $20$ $\mu \textrm{m}$ for a fixed pressure pulse does not significantly affect the temporal evolution of the bubble radius (see Fig. 3$(a)$ of Ref. \cite{bremond2006interaction}). In order to understand the effect of the nuclei radii on the asymmetry in multibubble cavitation, we performed a parametric study for the case where a pair of nuclei are separated by a distance $d 
= 200$ $\mu \textrm{m}$ with $c_e = 333 \textrm{ m/s}$, while varying the initial radii $R_{0}$ in the range $(20,50)$ $\mu \textrm{m}$. The evolution of asymmetry $\mathcal{A}_y (t)$ and the maximum value of the asymmetry $\textrm{max}(\mathcal{A}_{y})$ are shown in Figs. \ref{fig:AsymSize}$(a)$ and \ref{fig:AsymSize}$(b)$, respectively. Similar to the bubble radius evolution in the case of a single bubble, the asymmetry $\mathcal{A}_y (t)$ varies slightly with the size of the nuclei; thus, its effect is secondary in comparison to the effect of the effective speed of sound $c_e$.
The maximum asymmetry $\textrm{max}(\mathcal{A}_{y})$ increases with decreasing nuclei size and saturates for smaller $R_{0}$. This behavior is likely due to the increased shielding effect caused by the neighboring bubbles, which becomes increasingly important for larger nuclei at a fixed inter-bubble distance $d$. If we extrapolate these results to $5$ $\mu \textrm{m}$ nuclei, the result would indicate that the previous values of $c_e$ are underestimated, implying that the asymmetry in the experiments could be explained by even smaller gas concentrations.


\section{\label{sec:multi}Multiple bubbles}
\begin{figure}[b!]
    \centering
    \includegraphics{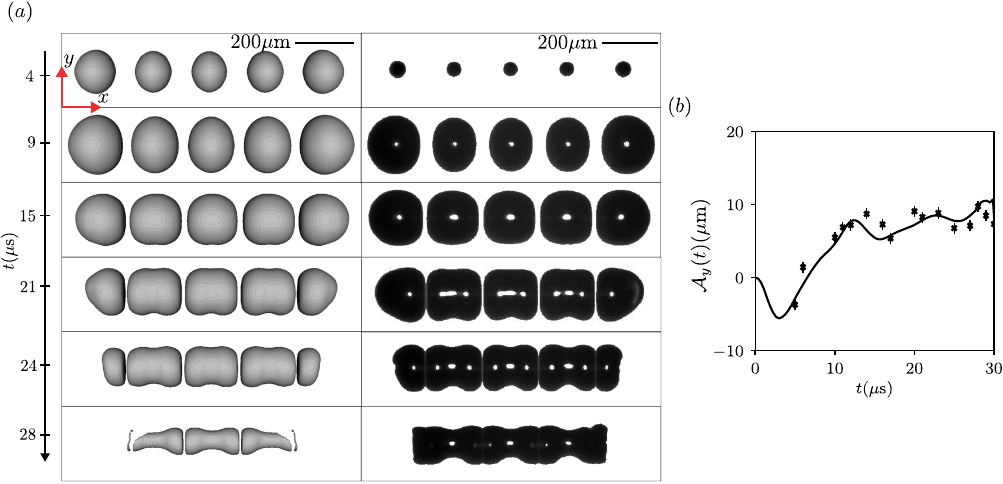}
    \caption{$(a)$ Top view of the five bubbles in a straight line configuration, as discussed in Ref. \cite{bremond2006interaction}. The numerical results are shown in the left panel, where the bubble expands from $20$ $\mu {\textrm m}$ hemispherical nuclei and the effective speed of sound is set to $667{\textrm m/s}$. The experimental bubble shapes are shown in the right panel. $(b)$ The evolution of the asymmetry parameter $\mathcal{A}_y(t)$ obtained from the numerical simulations (solid line) and the experiments (crosses). The error bar is equal to one pixel size (i.e., $1.53 \mu {\textrm m}$) in the experiment snapshots.}
    \label{fig:5Bub}
\end{figure}

We now consider the cases with five bubbles arranged in a line and with a cluster of 37 bubbles arranged in a hexagonal pattern \cite{bremond2006interaction,PhysRevLett.96.224501}. Like before, we investigate the influence of the effective speed of sound $c_e$ on the asymmetry by fixing $R_{0} = 20$ $\mu \textrm{m}$ and finding the values of $c_e$ that compare best with the experimental data.

In the case of the five bubbles, the best correspondence between the numerical and experimental results is obtained for $c_e = 667 \textrm{ m/s}$. In Fig. \ref{fig:5Bub}$(a)$, we show the top views of both experimental and numerical bubble shapes in the left and right panels, respectively. The snapshots exhibit very similar bubble dynamics, except at $t = 4$ $\mu \textrm{s}$, where the numerical bubbles are comparatively bigger due to the difference in the initial radii. Like for the bubble pair, we observe deformed bubble shapes during the collpase stage. We again use the asymmetry parameter $\mathcal{A}_y$ to quantify the asymmetry [Fig. \ref{fig:5Bub}$(b)$] which is shown to match well with the experiments for $c_e = 667 \textrm{ m/s}$.

\begin{figure}[htb!]
    \centering
    \includegraphics{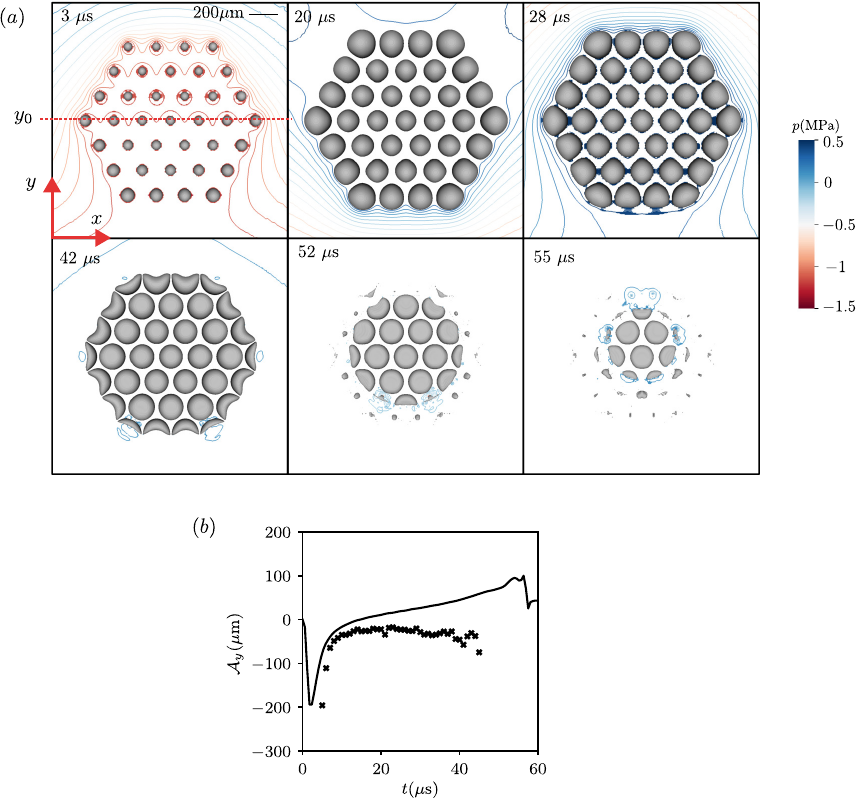}
    \caption{(a) Top view of the cluster of 37 bubbles in a hexagonal arrangement expanding from $20$ $\mu {\textrm m}$ hemispherical nuclei. The effective speed of sound in the liquid is $1480\textrm{ m/s}$. Isobars are shown with a colormap. $(b)$ The evolution of the asymmetry parameter $\mathcal{A}_y(t)$ obtained from the numerical simulations (solid line) and the experiments (crosses).}
    \label{fig:Hex}
\end{figure}

Finally, we investigate the most challenging problem of a cluster of 37 bubbles in a hexagonal pattern, for which we repeat the same procedure of varying $c_e$ for hemispherical nuclei with radii of $20$ $\mu \textrm{m}$ in order to reproduce the experimental measurements. Remarkably, the largest admissible value of $c_e = 1480 \textrm{ m/s}$ is better at reproducing the experimental results than the lower values of $c_e$. However, the numerical results show good resemblance only during the expansion stage (see Fig. \ref{fig:Hex}$(a)$ and \ref{fig:asymexpt}$(b)$). The bubble shapes during the collapse phase do not match well the experiments in which the collapse process is very chaotic and random [see Fig. \ref{fig:asymexpt}$(b)$]. In Fig. \ref{fig:Hex}$(b)$ we plot the asymmetry parameter $\mathcal{A}_y (t)$ as a function of time. Consistent with our previous observations, the value of $\mathcal{A}_y (t)$ computed from the numerical simulations changes sign during the collapse phase whereas that is not observed in the experimental data, where $\mathcal{A}_y$ remains negative. Furthermore, the agreement is not as satisfactory as in the previous cases. 

It is noteworthy that the effective speed of sound required to replicate the experiments with a small number of bubbles ($500$ m/s in the case of bubble pairs, $667$ m/s in the case of bubble quintets) was significantly lower than the speed of sound in clean water without bubbles ($1480$ m/s). The reason for this lies in the presence of the many small microbubbles which emerged from prior cavitation events. It is nearly impossible to quantitatively model these bubbles, due to the poor quantitative reproducibility. We expect the concentration of these microbubbles to be higher in the liquid volume surrounding the pits due to the continuous fragmentation of the bubbles over successive experiments and to decrease quickly with the distance. Assuming that (1) the number of bubbles produced by fragmentation is proportional to the number of bubbles $N_b$ and (2) the overall response of the bubble cluster is mainly influenced by the effective compressibility of the medium at a distance that scales with the characteristic length of the bubble cluster $L_c$, we find that the effective concentration of tiny bubbles in the medium surrounding the region with pits is inversely proportional to a power law of the number of bubbles $N_b$ with a positive power law exponent. This is probably the reason why, for systems with a large number of bubbles, the influence of tiny bubbles on the dynamics of the system is less important and approaches that of a pure liquid.

 \section{\label{sec:conclusion}Conclusions and outlook}
The asymmetry appearing due to the interaction of cavitation bubbles with the driving pressure pulse was studied in detail using three-dimensional direct numerical simulations of the compressible Navier-Stokes equations. The numerical results were compared to the experiments available in the literature. Our results demonstrate that a lower effective speed of sound in the medium better represents the experimental observations, particularly the asymmetry induced by the pressure pulse. This asymmetry is shown to scale with the pressure gradient imposed by the pressure pulse during the expansion phase. It also influences the direction of the jet generated during the last stages of the bubble collapse. As a physical justification for the change in the speed of sound, we postulate that the medium is polluted by small gas bubbles, originating from prior cavitation events. A simplified calculation based on the linear theory of wave propagation indeed suggests that a very small gas content can influence the results significantly. The effect of the radii of nuclei on the asymmetry remains secondary to the effect of the speed of sound. The same conclusions are valid for the setup with five bubbles, but the predicted effective speed of sound in the medium increases. On the other hand, for the cluster of 37 bubbles the numerical results do not agree well with the experimental results which are very chaotic. The effective speed of sound $c_e$ computed by the comparison of experimental and numerical data suggests that in this case direct bubble interactions play a more important role than the large-scale interactions of the bubble cluster with the pressure pulse. As a future outlook, one should try to systematically vary the bubble nuclei size or concentration and perform local measurements of the speed of sound to correlate with the present numerical results.

\begin{acknowledgments}
We greatly acknowledge the help of Dr. N. Bremond for digging out his the experimental data from 2005. This research is supported by the European Union (EU) under  MSCA-ITN Grant Agreement No. 813766, under the project named "Ultrasound Cavitation in sOft Matter (UCOM)\`. This work was also supported by a grant from the Swiss National Supercomputing Centre (CSCS) under project ID: TRUFLOW, s1136. 
\end{acknowledgments}

\appendix

\section{\label{App}Effect of the numerical slip length}

\begin{figure}[h]
    \centering
    \includegraphics[scale = 1.15]{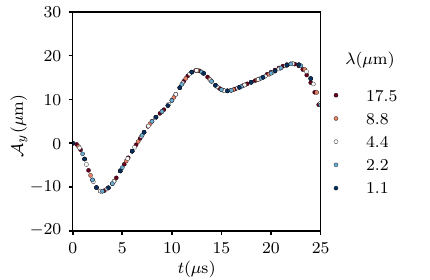}
    \caption{The time evolution of parameter $\mathcal{A}_y(t)$ is shown for varying values of a numerical slip length for the bubble pair case. The plot demonstrates the robustness of the results.}
    \label{fig:AsymSL}
\end{figure}

We also study the effect of the slip length on the asymmetry (see Fig. \ref{fig:AsymSL}) for the particular case of a pair of nuclei with radii $R_0 = 20$ $\mu \textrm{m}$ separated by $d = 200$ $\mu \textrm{m}$. The speed of sound is set to $c_e = 333 \textrm{ m/s}$, and the slip length $\lambda_{num}$ is varied in the range $(1.1,17.5)$ $\mu \textrm{m}$. As expected, the effect of the slip length and the contact line motion is local and does not influence the bubble shapes. This is an indirect verification that the results discussed in this article are not sensitive to the modeling of the contact line. Similar conclusions were drawn in Ref. \cite{saini2022dynamics} for jetting during the bubble collapse problem using the current method.


\bibliography{apssamp}

\end{document}